\def\aa{{A\&A}}
\def\aas{{A\&AS}}

\def\apj{{ApJ}}
\def\apjs{{ApJS}}

\def\mnras{{MNRAS}}

\documentclass[cup5b]{caps}
\usepackage{graphicx}
\usepackage{amssymb}
\usepackage{ociwsymp3e}  %Use this for contributed papers.

\newcommand{\Hb}{\ensuremath{{\rm H}\beta}}

\def\ga{\mathrel{\mathchoice {\vcenter{\offinterlineskip\halign{\hfil
$\displaystyle##$\hfil\cr>\cr\sim\cr}}}
{\vcenter{\offinterlineskip\halign{\hfil$\textstyle##$\hfil\cr
>\cr\sim\cr}}}
{\vcenter{\offinterlineskip\halign{\hfil$\scriptstyle##$\hfil\cr
>\cr\sim\cr}}}
{\vcenter{\offinterlineskip\halign{\hfil$\scriptscriptstyle##$\hfil\cr
>\cr\sim\cr}}}}}
\def\la{\mathrel{\mathchoice {\vcenter{\offinterlineskip\halign{\hfil
$\displaystyle##$\hfil\cr<\cr\sim\cr}}}
{\vcenter{\offinterlineskip\halign{\hfil$\textstyle##$\hfil\cr
<\cr\sim\cr}}}
{\vcenter{\offinterlineskip\halign{\hfil$\scriptstyle##$\hfil\cr
<\cr\sim\cr}}}
{\vcenter{\offinterlineskip\halign{\hfil$\scriptscriptstyle##$\hfil\cr
<\cr\sim\cr}}}}}

\HeadText{A. Fritz et al.}

\begin{document}

\pagenumbering{arabic}

\author[]{A. FRITZ$^{1,} \footnote{{\it E-mail: afritz@uni-sw.gwdg.de}}$,
B. L. ZIEGLER$^{1}$, R. G. BOWER$^{2}$, I. SMAIL$^{2}$, R. L. DAVIES$^{3}$
\\
(1) Universit\"atssternwarte G\"ottingen, Geismarlandstr. 11,
37083, G\"ottingen, Germany\\% E-mail: afritz@uni-sw.gwdg.de\\
(2) Department of Physics, University of Durham, Durham DH1\,3LE, UK\\
(3) University of Oxford, Astrophysics, Keble Road, Oxford, OX1 3RH, UK}

\chapter{Early-type Galaxies in the Cluster \\ Abell 2390 at z = 0.23}

\begin{abstract}
To examine the evolution of the early-type galaxy population in the rich
cluster Abell~2390 at $z=0.23$ we have gained spectroscopic data of 51
elliptical and lenticular galaxies with MOSCA at the 3.5~m telescope on
Calar Alto Observatory. This investigation spans both a broad range in
luminosity ($-19.3\geq M_{B}\geq-22.3$) and uses a wide field of
view of $10'\times 10'$, therefore the environmental dependence of
different formation scenarios can be analysed in detail as a function of
radius from the cluster centre. Here we present results on the surface
brightness modelling of galaxies where morphological and structural information
is available in the F814W filter aboard the {\it Hubble Space Telescope (HST)}
and investigate for this subsample the evolution of the Fundamental Plane.
\end{abstract}

\section{Introduction}

A key question of early-type galaxy evolution is when and within what timescales
their stellar populations have been formed. In models based on a monolithic
collapse a burst of star formation at high redshift ($z_{{\rm f}}\ga2$) is
followed by a passive evolution of the stellar populations.
This is in contrast to the formation scenario of hierachical cluster models,
presuming longer assembly timescales for the more massive galaxies and
therefore resulting in somewhat younger mean ages.

In order to probe the evolution of galaxies, it it necessary to explore both
the morphological evolution as well as the evolution of the luminosities and
the mass-to-light ($M/L$) ratios of the galaxies. This can be best achieved
with the Fundamental Plane (FP).
In a three dimensional parameter space, defined by three parameters,
the effective radius $R_{{\rm e}}$, effective surface brightness $\mu_{{\rm e}}$
and velocity dispersion $\sigma$, the FP establishes a tight correlation
(Dressler et al. 1987, Djorgovski \& Davis 1987). Projections of this plane
are the the Faber-Jackson relation (FJR),
luminosity $L$ {\it vs.} $\sigma$ relation, (Faber \& Jackson 1976) and the
Kormendy relation (KR), a correlation between $R_{{\rm e}}$ {\it vs.}
$\mu_{{\rm e}}$ (Kormendy 1977).

To overcome selection problems and possible limitations due to a small number
of the more luminous galaxies of previous spectroscopic samples, we focus in
this investigation of the cluster Abell~2390 on a large number of objects
($N=51$), spanning a wide range in luminosity $23.3>B>21.4$,
$-19.2>M_{B}>-22.4$ and a wide field of view of $\sim 10'\times 10'$
($2.5\times 2.5$~Mpc), analogous to the study of Abell~2218 by
Ziegler et al. (2001). 
Furthermore the evolution of galaxies in age, metallicity and abundance ratios
is investigated by analysing absorption line strengths, such as H$\beta$,
Mg$_{b}$, Fe5270 and Fe5335 in comparison with the latest stellar
population models. The determination of velocity dispersions of our
galaxies makes it possible not only to study the Faber-Jackson and Mg$-\sigma$
relations but also these relations in respect to the slope, scatter and
zeropoint. With this large sample, it is possible to explore variations in
early-type galaxy evolution not only in the dense core of a rich cluster but
also in a less dense environment by incorporating galaxies at larger radii
from the cluster centre. In addition different sub-populations (like E, S0,
E+A) can be analysed with statistically significant reliability.

Through this article we adopt a cosmological model with a deceleration
parameter of $q_{0}=0.1$, a Hubble constant of
$H_{0}$ = 65\,km\,s$^{-1}$\,Mpc$^{-1}$ and a cosmological constant of
$\Omega_{\Lambda}=0$.

\section{Observations and Sample Selection}
The large spectroscopic sample of this study comprises a total of 63 spectra
of 51 different early-type galaxies gained using the MOSCA
spectrograph at the 3.5~m telescope at Calar Alto Observatory in Spain during
two observing runs (Sept. 1999 and July 2000). 
The spectral resolution in the wavelength range
$5900\la\lambda\la 6400$~Angstr\"om (around the \Hb\ and Mg$_{b}$ lines) was
5.5~Angstr\"om FWHM, corresponding to $\sigma_{\rm inst}\sim 100$~km~s$^{-1}$. 
Values for the $S/N$ vary between 9.6 and 79.8 with an average value of
$S/N\sim 41$.

The objects were selected on the basis of ground-based Gunn $i$-band images
(500~sec) obtained with the Palomar 5~m Hale telescope. Additional imaging
data from Mt. Palomar is available in the $U$ (3000~sec) and $B$ (500~sec)
filter bands and the WFPC2 camera onboard the
{\it Hubble Space Telescope (HST)} observed A\,2390 in the F555W and F814W
filter (10800~sec each).

\section{Reduction and Analysis}
\begin{figure*}[t]
\vspace*{0.3cm}
\centerline{%
\begin{tabular}{c@{\hspace{0.3cm}}c}
\includegraphics[width=0.49\textwidth]{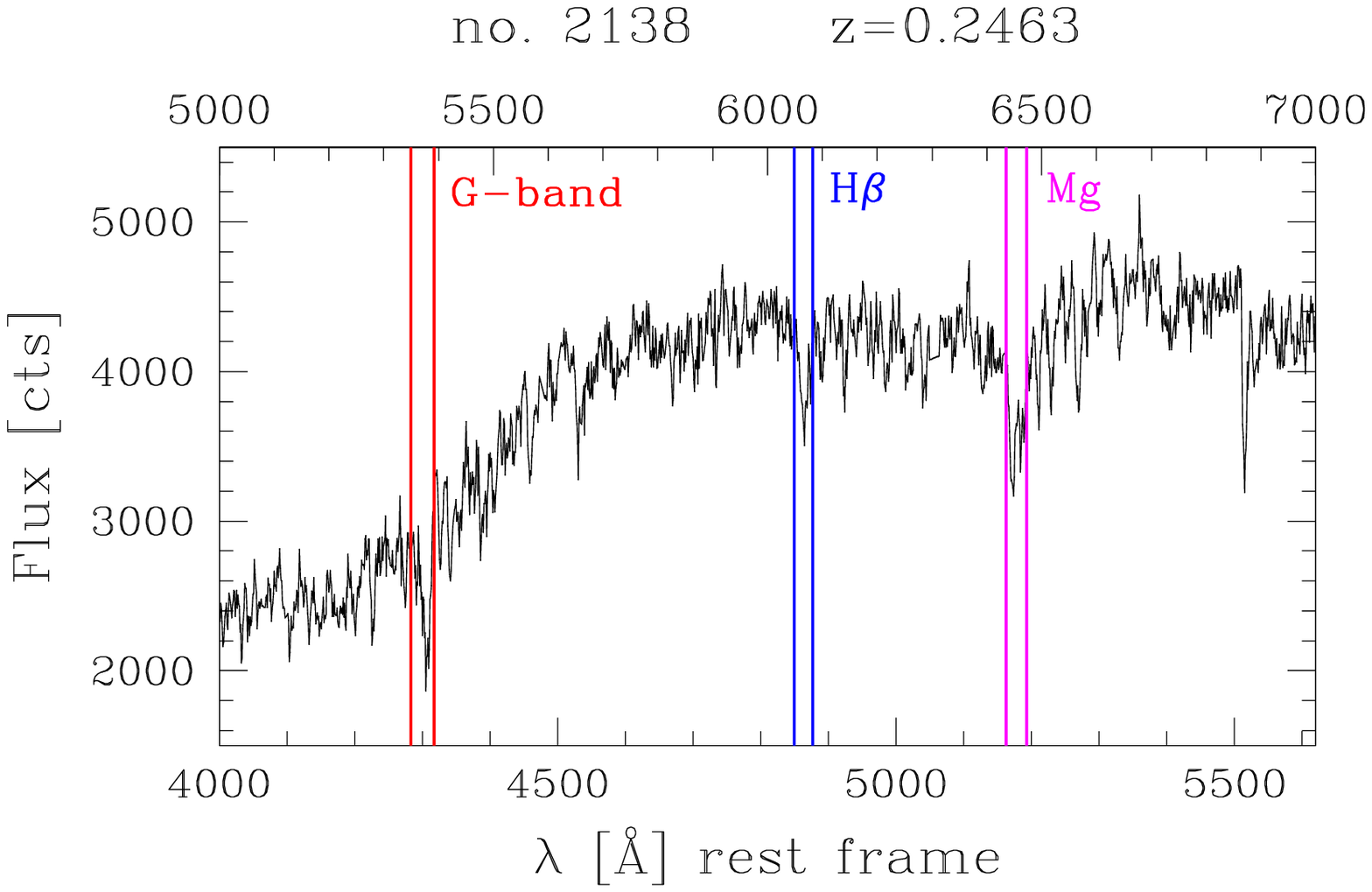} &
\includegraphics[width=0.49\textwidth]{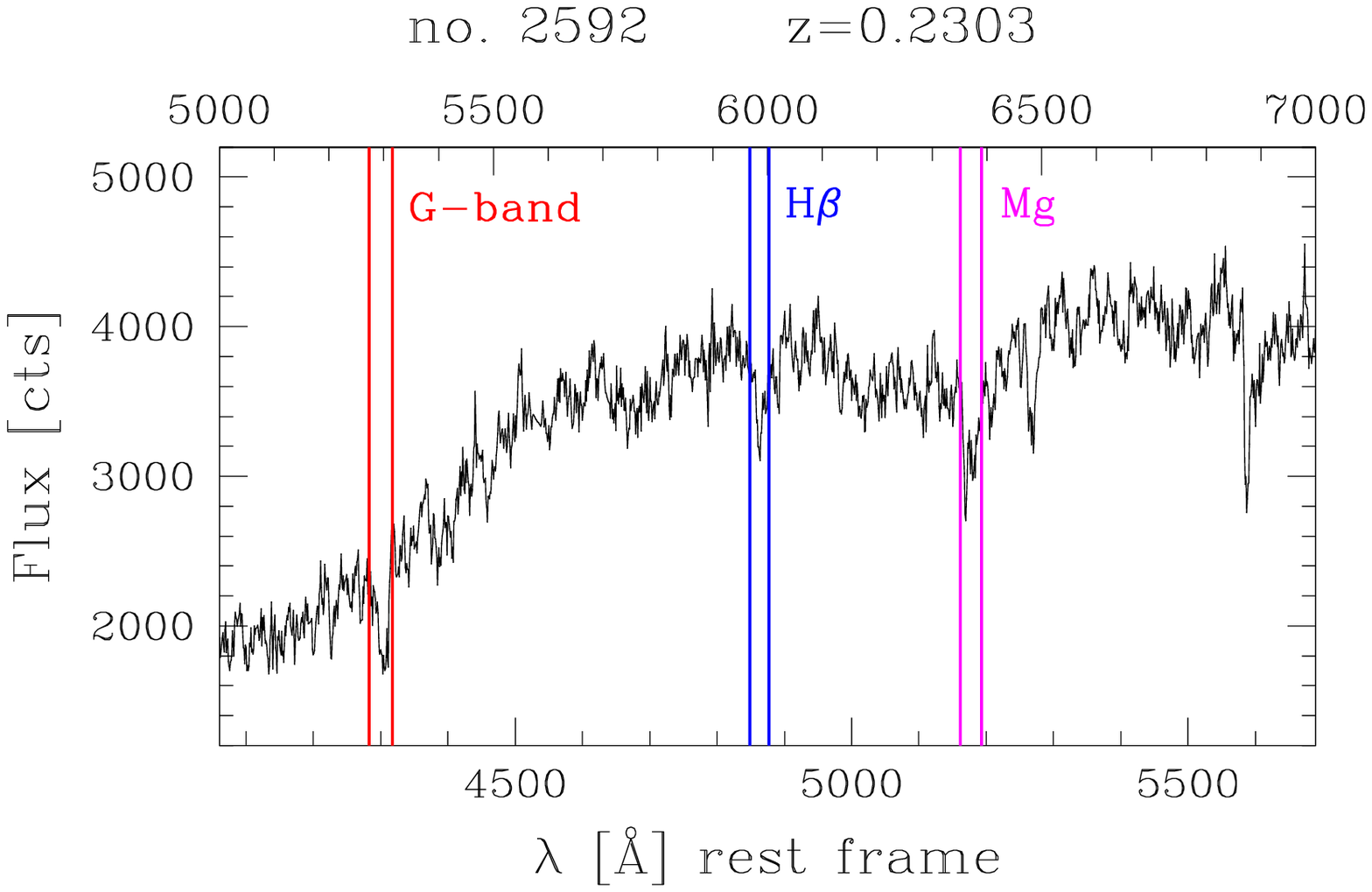} \\
\end{tabular}}
\caption{{\bf Upper panel:} $I_{814}$ Surface Brightness profiles
of the two objects \#\,2138 (S0) and \#\,2592 (E).
The total $I_{814}$ magnitude (ZP and extinction A$_{{\rm F814W}}$ corrected)
is plotted against the radius $R^{1/4}$ (in arcsec). Filled squares show
the observed profile, lines the different
best models for the bulge (de Vaucouleurs law), disk (exponential law)
component fit and for a combination of bluge and disk components. The arrow
indicates the position of the effective radius $R_{e}$. Different structural
parameters are shown (see text for details).
The inset gives the residuals $\Delta \mu_{I}$ between the observed profile and
the modelled fit as a function of $R^{1/4}$.
{\bf Lower panel:} Spectra (not flux-calibrated)
of the corresponding galaxies with determined redshift. The lower x-axis
represents the rest frame wavelengths, the upper one the observed wavelengths
(both in Angstr\"om). The ordinate gives the flux in counts
(1~ADU=1.1~$e^{-1}$). Prominent absorption features are marked.}
\label{sperestfr}
\end{figure*}

All spectra were reduced using standard reduction techniques implemented within
MIDAS and IRAF. Examples of final 1-dimensional spectra in rest frame
wavelengths are plotted in Fig.~\ref{sperestfr}. The velocity dispersions were
evaluated with the Fourier Correlation Quotient method as described
in Bender (1990).

Absolute magnitudes for all galaxies were calculated from our ground-based
$UB\,i$ photometry. Using SExtractor (Bertin \& Arnouts 1996), object positions
were determined and performing aperture photometry apparent magnitudes were
measured. The total Gunn $i$ magnitudes were transformed to rest frame $B$ and
rest frame Gunn $r$ magnitudes with typical k-corrections of
${m}_{B}\approx 1.2$ and ${m}_{r}\approx 0.34$, for HST objects according
to their morphology.

\section{HST Photometry}
The surface brightness models for our galaxies were constructed using our
F814W image. As pointed out by Ziegler et al. (1999), an exposure
time of $\sim 10$~ksec is deep enough to determine structural parameters down
to $M_{B\,{\rm rest}}\sim 23$ mag.

\subsection{Structual Parameter Analysis}
Structural properties were determined by fitting the surface brightness
profile as determined from the HST F814W images with an $R^{1/4}$ and an
exponential law profile, both separately and in combination
(Saglia et al. 1997). For the bulge profile a special form of the S\'{e}rsic
profile (S\'{e}rsic 1968), the classical de Vaucouleurs profile,
in the following form was applied:
\begin{equation}
\Sigma(R) = \Sigma_{e}\ {\rm exp}\ \{ -7.67\,[(R/R_{e})^{1/4}-1] \}
\end{equation}
$\Sigma(R)$ is the surface brightness at $R$ along the semimajor axis and
$\Sigma_{e}$ is the effective surface brightness.
The disk profile is well represented by an exponential law, defined as:
\begin{equation}
\Sigma(R) = \Sigma_{0}\ {\rm exp}\ \{-(R/h)\}
\end{equation}
where $\Sigma_{0}$ is the (face-on) central surface brightness for the disk and
$h$ the exponential disk scale length.
In total structural parameters could be determined for 14 galaxies out of 15
for which we also have obtained spectra. In Fig.~\ref{sperestfr} examples
of the surface brightness profile fitting are shown. Values for the total
magnitude $I_{tot}$, the central surface brightness $\mu_{0}$, the effective
radius $R_{e}$ and the effective radius of the bulge $R_{e,b}$, and the
disk-to-bulge ratio $D/B$ are listed. The model for the S0 galaxy \#\,2138
shows the existence of a small disk component at $R^{1/4}\sim 1.25$
($R\approx 2.44''$). In the case of the elliptical (E) galaxy \#\,2592 no
disk component is seen. Both galaxies are well reproduced by a combination of
a de Vaucouleurs profile and a model for an underlying disk component.

\section{The Fundamental Plane at intermediate Redshift}
\begin{figure*}[t]
  \centering
  \includegraphics[width=1.0\textwidth,angle=0]{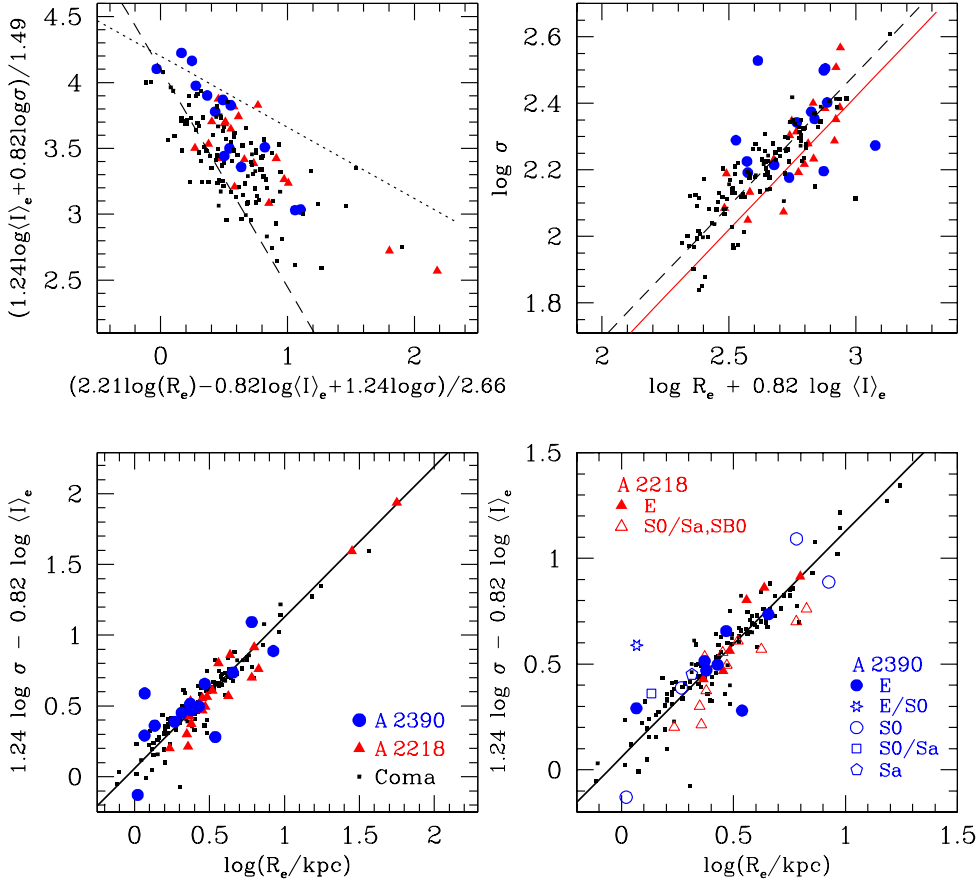} 
  \caption{Fundamental Plane for A\,2390 in rest frame Gunn $r$. Circles
  represent our A\,2390 objects, triangles the early-type galaxies of
  A\,2218 ($z=0.18$), small squares the Coma galaxies of JFK96 and J99,
  respectively.
  {\it Upper panel, left:} Face on FP. The dotted line indicates the so-called
  exclusion zone for nearby galaxies (Bender et al. 1992) and the dashed line
  the luminosity limit for the completeness of the Coma sample
  $M_{r\,{\rm T}}=-20.75$~mag. {\it Upper panel, right:} FP edge-on, along
  the short axis. The dashed line marks the relation for the local Coma sample,
  following JFK96, the solid line the fit for the A\,2218 galaxies. Our
  galaxies are more spread in respect to Coma but show a similar behaviour as
  the A\,2218 galaxies. 
  {\it Lower panel:} Edge-on FP. The solid line represents the fit for the local
  Coma objects. On the left side a zoom of the edge-on FP with a separation
  into different morphologies of the A\,2390 and A\,2218 galaxies is shown.
  Compared to the local Coma sample we find an FP evolution of
  $\overline{m}_{r}=0.12\pm 0.19$~mag.}
  \label{FPr}
\end{figure*}

In Fig.~\ref{FPr} the Fundamental Plane (FP) for A\,2390 in rest frame Gunn $r$
is illustrated. The figure also shows the FP for the local Coma samples of
J{\o}rgensen et al. (1996) (JFK96), J{\o}rgensen (1999) (J99) and for the
intermediate redshift cluster of A\,2218 by Ziegler et al. (2001) at $z=0.18$.
A morphological analysis of the HST galaxies revealed that our sample splits
nearly equally into elliptical and lenticular (S0) galaxies.
7 galaxies were classified as elliptical, 6 as lenticular and one as a
early-type spiral (Sa). Both families are uniformly distributed along the
Fundamental Plane (see Fig.~\ref{FPr}).
Compared to the sample of 115 Coma galaxies we deduce a modest luminosity
evolution of $\overline{m}_{r}=0.12\pm 0.19$~mag for the overall sample. This
result is in agreement with the predictions of the hierarchical
merging scenario for rich clusters (Kauffmann 1996).

\section{Conclusions}
We have investigated 51 members of the A\,2390 early-type galaxy population,
both over a broad range in luminosity ($-19.3\geq M_{B}\geq-22.3$) and a wide
field of view of $10'\times 10'$. Structural properties of the HST subsample
were analysed by combining an $R^{1/4}$ and an exponential law profile for
accurate surface brightness modelling.
From the Fundamental Plane for A\,2390 we find a mild evolution of
$\overline{m}_{r}=0.12\pm 0.19$~mag, which is in agreement with predictions of
hierachical models of cluster formation. \\

{\bf Acknowledgements:} AF and BLZ acknowledge financial support by the
Volkswagen Foundation (I/76\,520) and the Deutsche Forschungsgemeinschaft.
The Calar Alto staff is thanked for efficient observational support.

\newpage
% ################### Figure: Spectra of gal2138 and gal2592 ##################
\begin{figure*}[t]
\vspace*{0.3cm}
\centerline{%
\begin{tabular}{c@{\hspace{0.4cm}}c}
\includegraphics[width=0.5\textwidth]{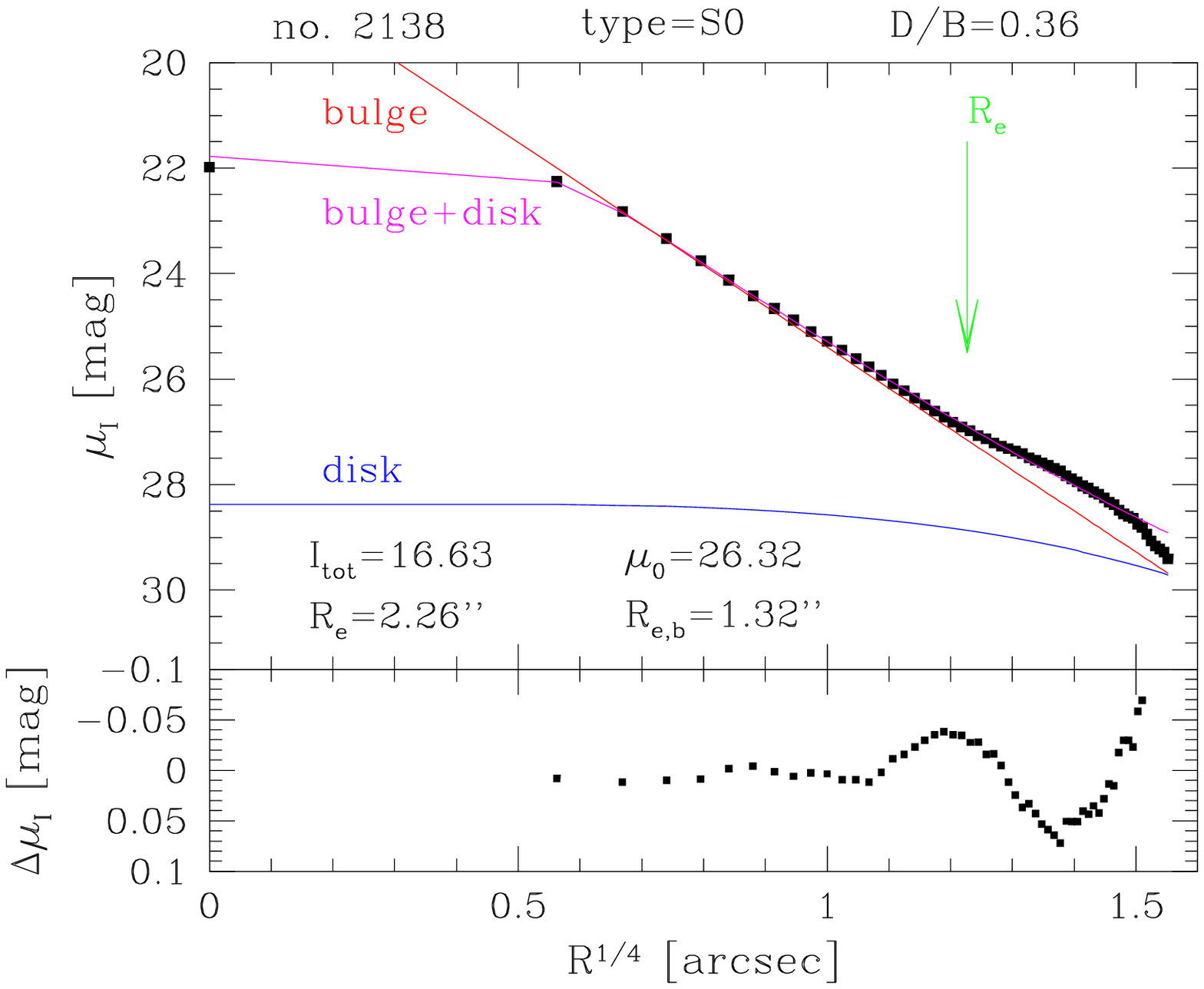} &
\includegraphics[width=0.5\textwidth]{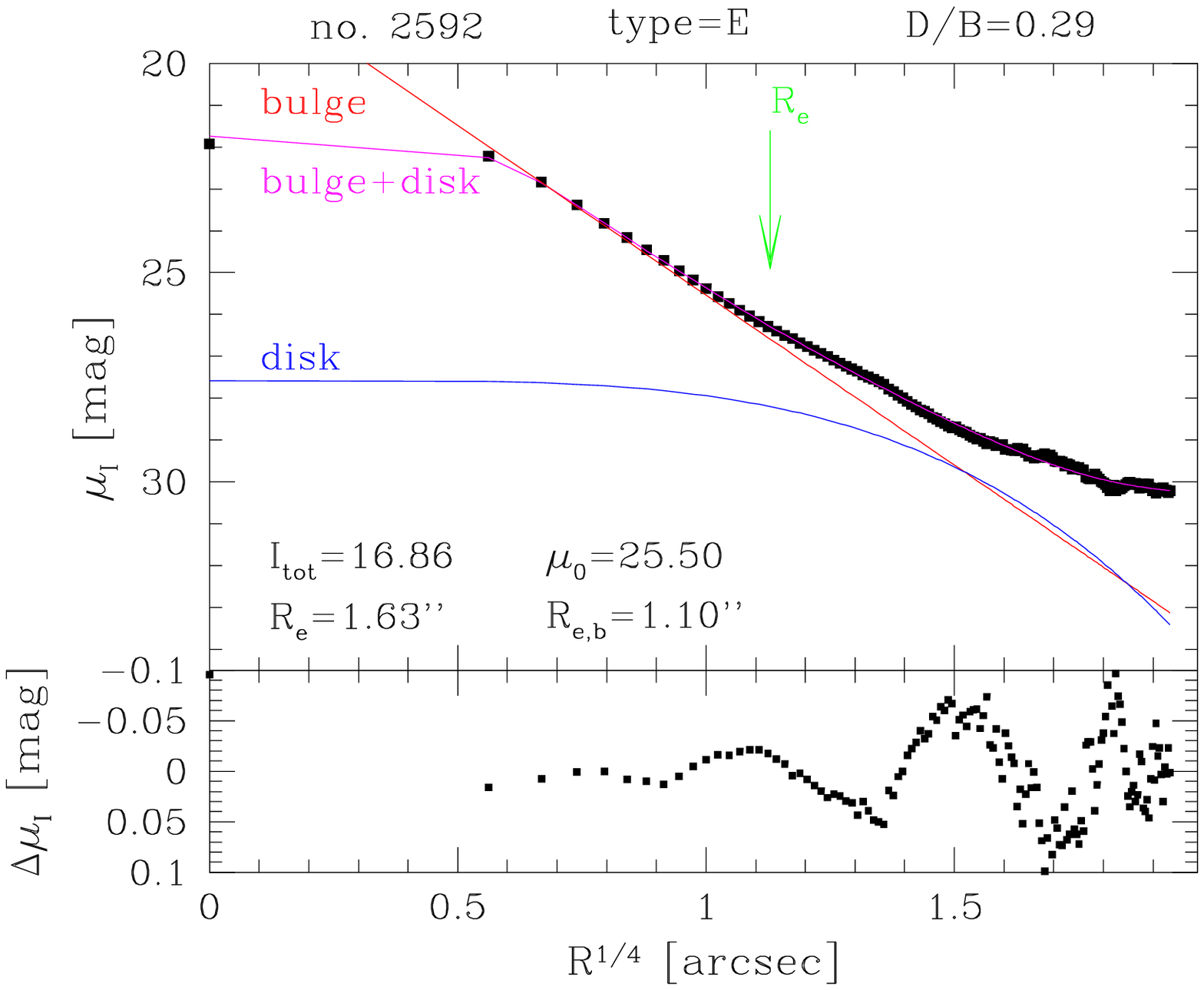} \\
\end{tabular}}
\end{figure*}
% ##############################################################################

\end{document}